# 3D simulations of the Archean Earth including photochemical haze profiles

**M. T. Mak[1], N. J. Mayne[1], D. E. Sergeev[1], J. Manners[2,1], J. K. Eager-Nash[1], G. Arney[3], E. Hébrard[1], K. Kohary[1]**

[1]Department of Physics and Astronomy, University of Exeter, Exeter, UK
[2]Met Office, Exeter, UK
[3]NASA Goddard Space Flight Center, Greenbelt, MD, USA

**Key Points:**

- A thin haze layer (at $CH_4/CO_2$ = 0.1) changes the abundance and location of water vapour and cloud formation, leading to global warming of ∼10.6 K.
- A thick haze layer (at $CH_4/CO_2$ > 0.1) reduces the amount of shortwave radiation reaching the deeper atmosphere, leading to global cooling.
- 3D models are crucial to correctly capture the warming and cooling effect of haze.

Corresponding author: Mei Ting Mak, M.Mak@exeter.ac.uk

**Abstract**

We present results from 3D simulations of the Archean Earth including a prescribed (non-interactive) spherical haze generated through a 1D photochemical model. Our simulations suggest that a thin haze layer, formed when $CH_4/CO_2 = 0.1$, leads to global warming of $\sim 10.6$ K due to the change of water vapour and cloud feedback, compared to the simulation without any haze. However, a thicker haze layer, formed when $CH_4/CO_2 > 0.1$, leads to global cooling of up to $\sim 65$ K as the scattering and absorption of short-wave radiation from the haze reduces the radiation from reaching the planetary surface. A thermal inversion is formed with a lower tropopause as the $CH_4/CO_2$ ratio increases. The haze reaches an optical threshold thickness when $CH_4/CO_2 \sim 0.175$ beyond which the atmospheric structure and the global surface temperature do not vary much.

## Plain Language Summary

The Archean era dates back to 2.5 to 4 billion years ago when the Sun was approximately 25%–20% less luminous than today. Many have suggested that the presence of methane ($CH_4$) and carbon dioxide ($CO_2$) could have kept the planet warm during the Archean, despite the fainter young Sun (Charnay et al., 2020). But laboratory data have shown that photochemical organic aerosols will start forming when the concentration of $CH_4/CO_2 \geq 0.1$ (Trainer et al., 2006). Here, we study the effect of these aerosols using a 1D photochemical model and a 3D climate model. We show that the presence of a thin aerosol layer will lead to warming of the planet due to the change of abundance and location of water vapour and clouds. A thick aerosol layer, on the other hand, will cool down the planet rapidly and make it uninhabitable. This work shows the importance of using a 3D model over a 1D model in studies of the climatic impact of organic hazes to capture the balance between the warming and cooling.

## 1 Introduction

Currently Earth is the only known example of an inhabited planet and life is believed to have been present very early in the Archean eon (which spanned $4 - 2.5$ billion years ago (Ga)) (Knoll & Nowak, 2017). The geological and atmospheric conditions of the Archean are an area of active research (Catling & Zahnle, 2020). Stellar models indicate that during the Archean the solar flux received by Earth was $\sim 25$–$20$ % less than the present day value $\sim 1361 \, W \, m^{-2}$ (Matthes et al., 2017). If the Archean Earth had the current atmospheric composition, it would have gone into complete glaciation (Kasting & Catling, 2003). However, sedimentary rock from at least 3.5 Ga has been found indicating different aqueous conditions, suggesting that the climate of Archean Earth was temperate with liquid water present (Benn et al., 2006; Feulner, 2012; Catling & Zahnle, 2020; Charnay et al., 2020). This discrepancy is termed the "Faint Young Sun (FYS) Paradox" (Sagan & Mullen, 1972). Solutions have been suggested in order to solve this paradox such as an increased ocean salinity, acting to facilitate oceanic heat transport (Olson et al., 2022), increased atmospheric abundances of greenhouse gases such as $NH_3$ (Sagan & Chyba, 1997), $C_2H_6$ (J. D. Haqq-Misra et al., 2008), carbon dioxide ($CO_2$) and methane ($CH_4$) (Charnay et al., 2013; Eager-Nash et al., 2023), and the changes of cloud structure due to this increased $CO_2$ concentration (Goldblatt et al., 2021).

$CH_4$ was indeed thought to be present in significant abundances during the Archean, leading to warming (J. D. Haqq-Misra et al., 2008). However, following on from the one-dimensional study of Byrne and Goldblatt (2015), Eager-Nash et al. (2023) performed 3D simulations of the Archean without the inclusion of photochemical haze and found surface warming reaches a maximum when the ratio of the partial pressure of $CH_4$ (p$CH_4$) to $CO_2$ (p$CO_2$) is approximately 0.1. Significant cooling was shown to occur at ratios more than 0.1 due to the strong shortwave absorption of $CH_4$. Before the Great Oxida-tion Event (GOE) which happened at 2.4 Ga, the Archean atmosphere was thought to be reducing due to the lack of oxygen in the atmosphere (unlike the present day atmosphere which is oxidising). This was inferred through the presence of sulfur isotope mass-independent fractionation (S-MIF) in Archean sedimentary minerals which indicates the production of elemental sulfur $S_8$ due to photochemistry in an anoxic environment (S-MIF would be nearly absent in an oxidising environment) (Farquhar et al., 2000; Zahnle et al., 2006; Claire et al., 2014; Catling & Zahnle, 2020). Methanogenic bacterial production of $CH_4$ was expected to be the major source of $CH_4$ in the atmosphere (Kharecha et al., 2005). Kharecha et al. (2005) have shown that Archean $CH_4$ atmospheric concentration could range from 100 parts per million by volume (ppmv) to 35,000 ppmv. An example $CH_4$ production pathway via methanogenisis is,

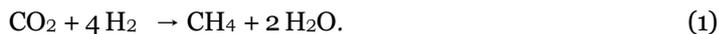

$$CO_2 + 4\,H_2 \;\rightarrow CH_4 + 2\,H_2O. \tag{1}$$

An anoxic atmosphere also increases the lifetime of $CH_4$ as it prevents it from being oxidised, through the net reaction,

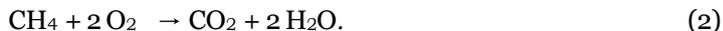

$$CH_4 + 2\,O_2 \;\rightarrow CO_2 + 2\,H_2O. \tag{2}$$

The signature of S-MIF from $SO_2$ photolysis also indicates the presence of species which absorb ultraviolet (UV) radiation, potentially a photochemically produced haze (Domagal-Goldman et al., 2008; Ueno et al., 2009; Zerkle et al., 2012; P. Liu et al., 2019). The Cassini/Huygens mission to Titan has allowed us to place constraints on $CH_4$ photochemistry and the properties of organic haze (Tomasko et al., 2008). Laboratory studies have also been carried out to study the formation of photochemical organic haze (Hasenkopf et al., 2011; Hicks et al., 2016). Even though the atmosphere of Titan is more reducing than that of the Archean Earth and their $CH_4$ photochemistry would be different, we do not have enough constraints on the properties of the Archean haze. Therefore applying Titan-haze analogues to the Archean is a good starting point to study the potential effect of haze on the Archean climate (Khare et al., 1984; Hasenkopf et al., 2010; Gavilan et al., 2017; Ugelow et al., 2018; He et al., 2022). Trainer et al. (2004, 2006) found that hydrocarbon haze starts to form when $CH_4/CO_2$ exceeds 0.1 from their experimental data on the photolysis of $N_2$, $CH_4$ and $CO_2$. This result has been studied in 1D models to explore the impact of haze on the Archean climate. For example, J. D. Haqq-Misra et al. (2008) finds a drastic cooling occurs when $CH_4/CO_2 = 0.1$. Zerkle et al. (2012) studied how the haze layer produced by different $CH_4/CO_2$ ratios changes the S cycle, and compared this with the S-MIF variations. They conclude that $SO_2$ photolysis might be the main source of S-MIF, with $CH_4/CO_2 < 0.1$ and $CH_4/CO_2 \approx 0.2$ being the stable regimes of their simulations. Zerkle et al. (2012) also derive an upper limit of $CH_4/CO_2 = 0.2$, above which haze production shuts down, or the entire Earth would have gone into glaciation, if the haze is being treated as fractal agglomerates (discussed in Section 4.2). Using the chemical network developed by Pavlov et al. (2001) and the laboratory data from Trainer et al. (2006), Arney et al. (2016) coupled the chemistry scheme in their 1D photochemistry model (Atmos) to their 1D radiative and climate model. By varying the $CH_4/CO_2$ ratio, they have found that at a $CH_4/CO_2 = 0.1$ the atmosphere including haze is radiatively indistinguishable from that of a clear sky simulation despite the antigreenhouse effect of the haze itself. As the $CH_4/CO_2$ ratio is increased, the haze can cool the globally averaged surface temperature by approximately 20 K. As the haze layer increases in thickness, the cooling reaches a plateau as it acts to shield $CH_4$ from further photolysis. However, as the study of Arney et al. (2016) uses a 1D model, the impacts of the global circulation, patchy clouds, and other effects are not taken into account, and their surface temperature may differ from 3D model results by a few K for haze-free atmospheres.

In this paper, non-interactive prescribed spherical photochemical haze is, for the first time, incorporated into a 3D General Circulation Model (GCM) simulation of the

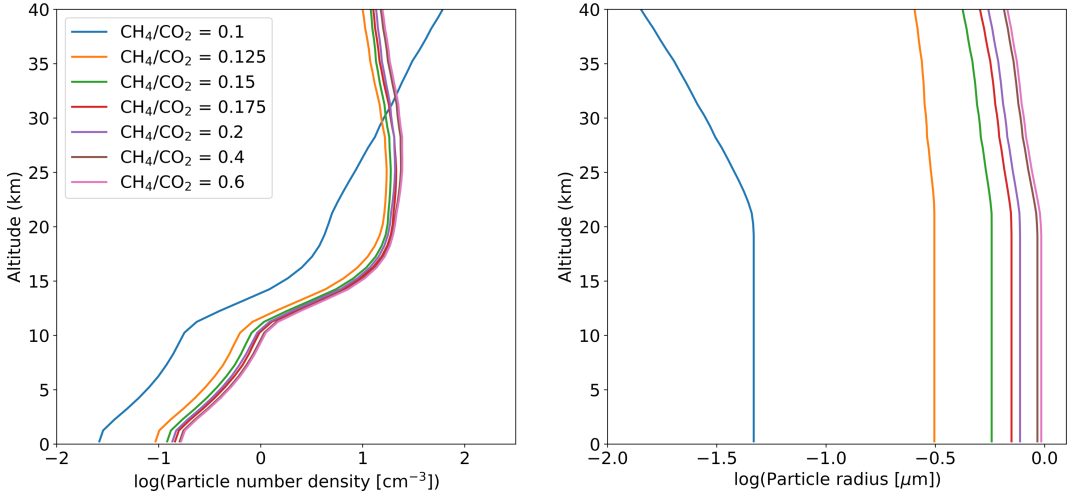

**Figure 1.** Particle number density in cm$^{-3}$ (left) and particle radius in $\mu$m (right) from Atmos, for each of the haze profiles.

Archean Earth atmosphere. For varying ratios of $CH_4/CO_2$, we obtain corresponding number densities and radii distribution for haze particles using the Atmos 1D photochemical model (as used by Arney et al., 2016). These profiles are then included in 3D simulations with the corresponding atmospheric $CH_4/CO_2$ ratio to explore the interaction with the global circulation. The GCM we use is the Met Office Unified Model (UM). The rest of this paper is laid out as follows. The configuration of the photochemical, radiative transfer and climate models are described in Section 2. The results are then presented in Section 3. In Section 4, we compare our results with those of previous 1D studies and discuss future directions of development. Finally, we present our conclusions in Section 5.

## 2 Models

We simulate the photochemical haze production in the Archean by running the 1D photochemical model Atmos, and prescribe the produced haze profiles within the 3D GCM the UM. The photochemical model is described in Section 2.1, followed by the radiative transfer model and climate model descriptions in Section 2.2 and 2.3, respectively.

### 2.1 Photochemical Model

We use the 1D photochemistry module from the Atmos model (available at https://github.com/VirtualPlanetaryLaboratory/atmos) to simulate photochemical haze formation in the Archean atmosphere. The Archean scheme is based on the model described in Arney et al. (2016), with the updates described in Teal et al. (2022) and Lincowski et al. (2018). In particular, these updates provide (1) an increased wavelength grid (750 bins compared to the previous 118 bin grid), (2) improved photolysis cross sections and quantum yields including for $H_2O$ as described in Ranjan et al. (2020), (3) an improved haze formation treatment where haze particles formed by each production path are grouped as a single population rather than in their own distinct population, and (4) new organic aerosol optical constants Gavilan et al. (2017). We adopt the boundary conditions used in Teal et al. (2022), with $CH_4$ and $CO_2$ mixing ratios described below.

The model includes 433 chemical reactions and 76 chemical species. Organic haze particles form through reactions:

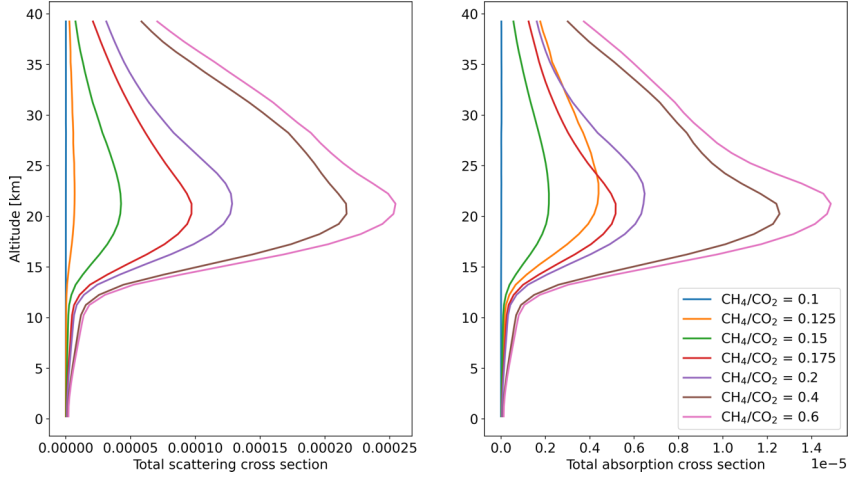

**Figure 2.** Total scattering (left) and absorption (right) cross section of each haze profile at wavelength 0.45 $\mu$m.

$$C_2H + C_2H_2 \;\rightarrow C_4H_2 + H, \tag{3}$$

$$C_2H + CH_2CCH_2 \;\rightarrow C_5H_4 + H. \tag{4}$$

Since the full chemical scheme that leads to haze formation is not well understood despite both laboratory and theoretical studies (Hallquist et al., 2009; Hicks et al., 2015), it is assumed that $C_4H_2$ and $C_5H_4$ condense directly to haze particles, with $C_4H_2$ being the dominating haze species.

Laboratory data from Trainer et al. (2004, 2006) shows that hydrocarbon haze starts forming when $CH_4/CO_2$ is approximately 0.1. Eager-Nash et al. (2023) has shown that surface warming reaches a maximum when $pCH_4/pCO_2$ is roughly 0.1 without haze. Beyond this ratio, cooling occurs at lower $pCO_2$, which then plateaus for higher $pCO_2$ due to its greenhouse effect. To study the cooling impact solely due to the haze, we have chosen the configuration of Eager-Nash et al. (2023), which also employs the same climate model that we use in this paper, with $pCO_2$ fixed at 3,000 Pa. In this case, Eager-Nash et al. (2023) has shown that increasing the $pCH_4$ does not lead to significant additional cooling due to the already high concentration of $CO_2$, which leads to a warmer atmosphere with more water vapour, resulting in a smaller change in shortwave absorption from $CH_4$ (see their discussion). Again note that their work did not include photochemical haze. Therefore, any significant decrease of the temperature in our experiments will be caused by the effect of the additional haze only. Keeping the atmosphere surface pressure constant at $10^5$ Pa in all cases, we calculate haze profiles using Atmos and the mean temperature-pressure profiles of Eager-Nash et al. (2023) for fixed $pCO_2$ (3,000 Pa), but for a range of $pCH_4$ values, such that $pCH_4$ = 300, 375, 450, 525, 600, 1,200 and 1,800 Pa, with ratios of $CH_4/CO_2$ = 0.1, 0.125, 0.15, 0.175, 0.2, 0.4 and 0.6, respectively.

In the photochemical model, the haze particles are treated as fractal particles, with a mass density of 0.64 g cm$^{-3}$ (Pavlov et al., 2001; Arney et al., 2016). Figure 1 shows the output particle number density (cm$^{-3}$) and particle radius ($\mu$m) from Atmos using the various $CH_4/CO_2$ ratios studied in this work. The radius used here is the radius of an equal mass spherical particle compared to the fractal particles. This radius is typically similar within an order of magnitude to the effective geometric radius of fractals generated in the photochemical model. When $CH_4/CO_2 = 0.1$, haze is produced with the smallest particle radius. All the other profiles share a similar number density distribution but the particle radius increases as the $CH_4/CO_2$ ratio increases. We assume the radius distribution in each layer to be log-normal with a geometric standard deviation of 1.5.

### 2.2 Radiative Transfer Model

We use the "Suite Of Community RAdiative Transfer codes based on Edwards and Slingo (1996)" (SOCRATES), to parametrise the optical properties of the haze. In SOCRATES, the haze particles are assumed to be spherical. Future work will update the model to include fractal particles. We apply Mie theory to calculate the scattering and absorption coefficients and the asymmetry factor of the haze. The optical properties are taken from Khare et al. (1984) and He et al. (2022). Khare et al. (1984) generate thin films of tholins by electric discharge of a mixture of 0.9 $N_2$ gas and 0.1 $CH_4$ gas by volume. Their wavelength range covers 0.027 $\mu$m to 920 $\mu$m. He et al. (2022) describes the updated refractivity of tholins using vacuum spectroscopy by mixing 5% of $CH_4$ in $N_2$ for wavelength ranges from 0.4 $\mu$m to 3.5 $\mu$m. We plot the total scattering and absorption cross section of each haze profile in Figure 2 at wavelength $\sim$0.45 $\mu$m of the peak intensity from the Archean solar spectrum. The total scattering (absorption) cross section is calculated to be the number density multiplied by the geometric cross section of the sphere ($\pi r^2$ where $r$ is the particle radius) and the scattering (absorption) efficiency factor. Figure 2 shows that the haze has the strongest radiative effect from the mid to upper atmosphere, and when $CH_4/CO_2 = 0.6$.

In order to run the Socrates two-stream radiative transfer code in the climate model, configuration files ("spectral files") are constructed that describe the wavelength breakdown into bands for the shortwave and longwave parts of the spectrum (Amundsen et al., 2014; Lines et al., 2018; Manners et al., 2022). The extinction, scattering and asymmetry coefficients for the haze are calculated for these bands, while the prescribed mass mixing ratios are specified directly within the climate model at runtime.

Along with the spherical haze optical properties, the spectral files also describe optical properties for the other radiatively active species. The shortwave part represents the solar radiation covering the wavelength range of 0.2 $\mu$m - 20 $\mu$m and is separated into 43 bands. Similar to the photochemical model (Section 2.1), we use the solar spectrum of the Sun at 2.9 Ga (Claire et al., 2012). The longwave part represents the thermal radiation from the planet and atmosphere. It covers the wavelength range of 3.33 $\mu$m - 10 mm and is divided into 17 bands. Within each band gaseous absorption is treated using the correlated-$k$ method. $k$-terms are generated using absorption line lists from HITRAN 2012 (Rothman et al., 2013). Collision-induced absorption from the gases are included, covering $N_2$-$CH_4$, $N_2$-$N_2$, $CO_2$-$CO_2$ from HITRAN (Karman et al., 2019), and $CH_4$-$CO_2$ (Turbet et al., 2020).

### 2.3 Climate Model

We perform climate simulations of the Archean Earth including haze using a 3D GCM, the Unified Model (UM), developed by the Met Office and used extensively for the study of modern Earth (Walters et al., 2019; Andrews et al., 2020). The UM has also been adapted for the study of a wide range of planets from Mars (McCulloch et al., 2023), to exoplanets (planets orbiting stars other than the Sun) that are terrestrial (Boutle et al., 2020; Sergeev et al., 2022; Ridgway et al., 2023) and gas giant (Mayne et al., 2019;

Christie et al., 2021; Zamyatina et al., 2023) exoplanets. The UM has also previously been used to study the Archean, focused on haze-free simulations of the impact of varying $CH_4/CO_2$ ratios on the surface temperatures (Eager-Nash et al., 2023), a study which we build on here. The dynamical core of the UM, ENDGame (Even Newer Dynamics for General atmospheric modelling of the environment), solves the non-hydrostatic, deep-atmosphere equations of motion in the atmosphere with varying gravity (see for discussion Wood et al., 2014).

This study adopts the setup of Eager-Nash et al. (2023), with their model framework described in their Section 2. The Archean Earth is treated as an aquaplanet with the planet surface being a single homogeneous and flat layer (Frierson et al., 2006), neglecting oceanic and surface heat transport. Including a dynamic ocean model requires setting the land configuration and bathymetry, neither of which are well constrained for the Archean. This would also add complexity which will inhibit our understanding of the impacts of haze, which is the focus of this work. The emissivity of the surface is fixed at 0.985. The sea liquid/ice albedo is wavelength dependent. Above 273.15 K, the albedo of liquid water is fixed at 0.07. Below 273.15K, the albedo becomes temperature-dependent as well using the HIRHAM parameterisation from J. Liu et al. (2007), with the maximum albedo fixed at 0.8 for wavelengths shorter than 0.5 $\mu m$, and 0.05 for wavelengths longer than that. This ice scheme underestimates the ice-albedo feedback but allows us to compensate for the missing dynamics of sea ice and oceanic heat transport (see discussion in Eager-Nash et al., 2023). Clouds are included using the PC2 scheme (Wilson et al., 2008). We only consider $N_2$, $CO_2$ and $CH_4$ in the atmosphere. As mentioned in Section 2.1, the surface pressure is fixed at $10^5$ Pa and the surface partial pressure of $CO_2$ is fixed at 3,000 Pa.

Earth's day length is fixed at 17 hours (Williams, 2000; Bartlett & Stevenson, 2016) and the eccentricity is fixed at 0 (Charnay et al., 2013; Wolf & Toon, 2014). The Earth's current obliquity of 23.4 ̊is used. Horizontal grid spacings of 2.5 ̊in longitude and 2 ̊in latitude are used. The vertical grid is quadratically stretched into 38 layers with the top level (top-of-atmosphere) at 39.25 km. This allows a higher resolution near the surface. The haze layers are fixed in time and space i.e. prescribed based on the results of the photochemical model. Our main focus in this work is to explore the impact of the haze on the thermal and dynamic structure of the atmosphere in an idealised way, before we build to a fully-interactive haze in future work. Once a given simulation reaches a steady state and the surface temperature fluctuation is less than 0.5 K, it is run for another standard length of 20 years, with the results then temporally averaged (mean) and analysed. In other words, all our data presented are based on 20 year mean climate state. We have also run the simulations without haze, but with the same $CH_4/CO_2$ ratios. For brevity, our simulations without haze are labelled "noAER" and those including haze "AER". For easier interpretation, haze is also labelled as "thin" when the $CH_4/CO_2$ ratio is small with a small extinction cross section and optical thickness, and "thick" when the $CH_4/CO_2$ ratio is large with a large extinction cross section and optical thickness.

## 3 Results

In the following sections, we start by presenting the vertical temperature structure of our simulations. We then separate the results into two sections, one for the simulation where haze acts to warm the surface, and another where the haze acts to cool the surface.

### 3.1 Vertical Temperature Structure

We find that a in in a non-interactive spherical haze setup, thin layer of haze can lead to global warming while a thick layer of haze can lead to global cooling. Figure 3 shows the global mean vertical temperature structure of the noAER and AER cases. For the noAER cases on the left, each case exhibits the same vertical temperature structure in the lower atmosphere. At higher altitudes, the cases with larger $CH_4/CO_2$ ratios show higher temperatures. This is due to the increase of shortwave absorption from the increase of $pCH_4$ (see Eager-Nash et al., 2023, for details). On the right, the vertical temperature profiles for the AER case are compared with the noAER case when $CH_4/CO_2$ = 0.1. When $CH_4/CO_2 = 0.1$, the inclusion of haze leads to a warmer atmosphere, especially above $\sim 17$ km. The haze layer is thin enough that shortwave radiation can be absorbed throughout the atmosphere, without developing a temperature inversion. When $CH_4/CO_2 > 0.1$, the lower atmosphere shows a significant cooling and the upper atmosphere exhibits very strong warming and becomes almost isothermal. We define the tropopause as the region in which more than two consecutive intervals have a lapse rate of less than $2\,K\,km^{-1}$, similar to the definition taken from World Meteorological Organisation (1957). Table 1 indicates the altitude of the tropopause in each AER case when $CH_4/CO_2 > 0.1$. As seen also in Figure 3, the tropopause is lower for a higher $CH_4/CO_2$ ratio. This is due to the increasing contrast of shortwave absorption in the upper and lower atmosphere as the haze layer becomes thicker and a reduction of radiation reaching the surface. Therefore resulting in the temperature inversion occurring lower in the atmosphere.

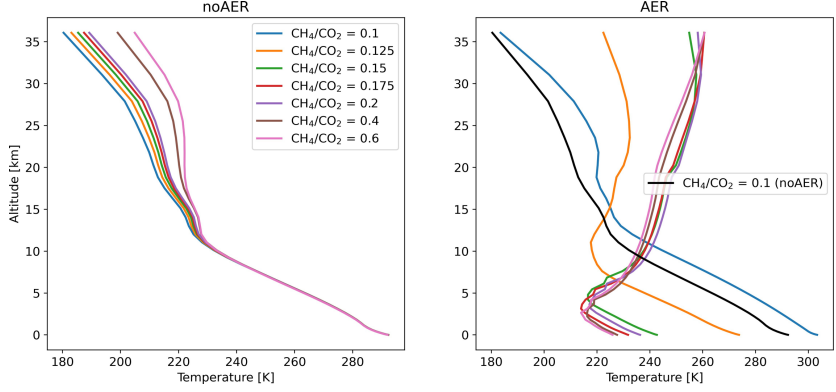

**Figure 3.** Vertical temperature profiles of noAER (left) and AER simulations (right). The noAER cases all have an identical structure in the lower atmosphere but a warmed upper atmosphere when the a larger $CH_4/CO_2$ ratio increases. For AER cases when $CH_4/CO_2 = 0.1$, the atmosphere is warmer than the noAER case but as the ratio increases, the heating in the upper atmosphere increases and the cooling in the lower atmosphere also increases. This results in a temperature inversion and the lowering of the tropopause.

**Table 1.** Altitude of tropopause [km] for AER cases when $CH_4/CO_2 > 0.1$.

| $CH_4/CO_2$ | 0.125 | 0.15 | 0.175 | 0.2 | 0.4 | 0.6 |
|---|---|---|---|---|---|---|
| Altitude [km] | 7.5 | 3.5 | 2.5 | 3 | 2 | 2 |

Figure 4 shows the global mean surface temperature of the noAER and AER cases. For the noAER cases, the surface temperature remains constant at $\sim$292.9 K as the $CH_4/CO_2$ ratio increases (as in Eager-Nash et al., 2023). For the AER cases, the surface temperature goes up to $\sim$303.5 K when $CH_4/CO_2 = 0.1$. As the ratio increases, the temperature drops rapidly to $\sim$226.8 K when $CH_4/CO_2 = 0.6$. This matches what is shown from the vertical temperature profiles in Figure 3, with global warming happening when $CH_4/CO_2$

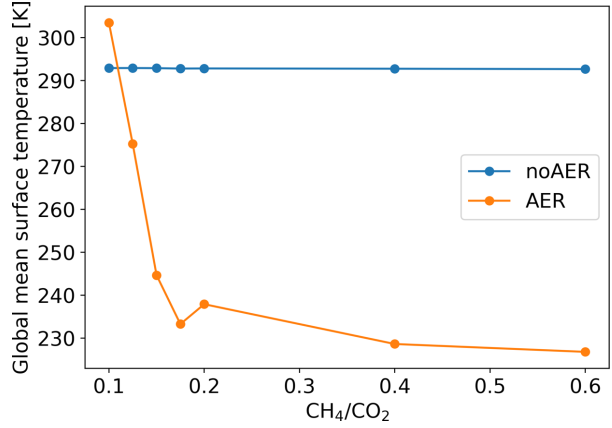

**Figure 4.** Global mean surface temperature of the noAER and AER cases.

= 0.1 and global cooling occurring when $CH_4/CO_2 > 0.1$. We observe a slight drop of the temperature in the AER case when $CH_4/CO_2 = 0.175$ compared to the general trend. This is likely due to the balance of haze scattering and absorption strength in the atmosphere which reduce the amount of shortwave flux absorbed by the surface, compared with when $CH_4/CO_2 = 0.2$. However, this does not affect the general trend of the drop in surface temperature with increasing $CH_4/CO_2$.

We also explore the habitable condition of our simulations of Earth during the Archean, by studying whether liquid water can be sustained on the surface. Figure 5 shows the zonal mean surface temperature of the AER cases. Here we define the habitability threshold to be 273 K (water freezing point), which is also indicated in the figure. We only consider the cases where cooling occurs, which is when $CH_4/CO_2 > 0.1$. The case when $CH_4/CO_2 = 0.1$ is not included as its minimum surface temperature is above 273 K. The maximum zonal mean surface temperature goes below 273 K when $CH_4/CO_2 \geq 0.15$. We therefore conclude that only the simulations with $CH_4/CO_2 < 0.15$ in our work here are considered habitable.

To summarise, the haze profile for $CH_4/CO_2 = 0.1$ is exhibiting a warming effect on the atmosphere while the others are showing a cooling effect. In the following subsections, we examine the reasons behind warming at $CH_4/CO_2 = 0.1$, and those behind cooling at $CH_4/CO_2 > 0.1$, separately.

### 3.2 $CH_4/CO_2 = 0.1$: Haze Warming Effect

Figure 6 shows the zonal mean air temperature difference between the cases with and without haze (AER - noAER) when $CH_4/CO_2 = 0.1$. The overlapping contours are indicating the corresponding difference of zonal mean shortwave heating rate. The vertical coordinate for Figure 6, $\sigma$, is the air pressure divided by the pressure at the surface. Almost the entire atmosphere is warmer for the haze case. The maximum warming occurs at latitudes of $\sim 40°$. Only the high-altitude ($\sigma \sim 0.1$) polar regions are cooler. To understand the warming due to the presence of haze, we first examine the strength of scattering and absorption by the haze itself. The heating rate in Figure 6 is the largest in the presence of haze at the top of the atmosphere and as it goes towards the surface, the heating rate differs from $\sim 2.5$ to $0.06 \, \mathrm{K \, day^{-1}}$, but is not significant enough to lead to the increase of global mean surface temperature of 10.6 K. Figure 7, left panel, shows the differences in the global mean clear-sky shortwave flux between the cases with and without haze (AER - noAER). Clear-sky flux means that the flux takes into account the radiative effect of hazes and water vapour, but not clouds from the PC2 scheme. From Figure 7 less shortwave flux is scattered back upwards in the AER case, with the decrease of flux being less than 1 W m⁻². This indicates more shortwave flux is absorbed by the atmosphere in the presence of a thin haze, and slightly less solar radiation is entering the lower atmosphere as well due to haze absorption, as seen from the increasing difference of downward shortwave flux as $\sigma$ increases. The right panel of Figure 7 shows the differences in the global mean clear-sky longwave flux. Both the upward and downward longwave flux exhibit an increase due to the haze and water vapour absorbing shortwave radiation and re-emitting longwave radiation.

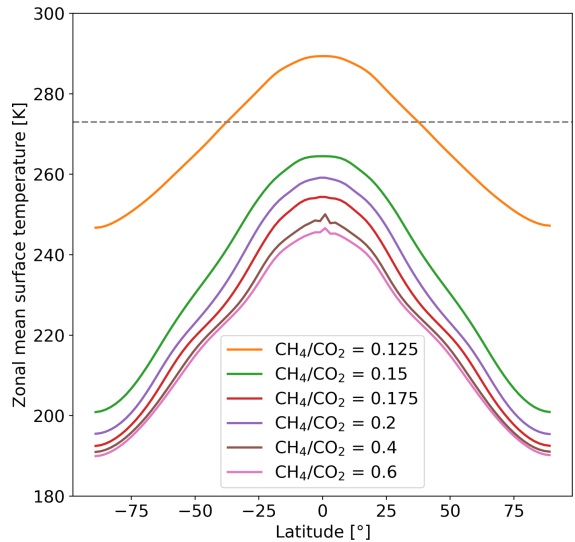

**Figure 5.** Zonal mean surface temperature of AER cases when $CH_4/CO_2 > 0.1$. The boundary of 273 K is marked as the habitability threshold.

The longwave radiation re-emitted by the haze will heat up the surface. This leads to an increased surface evaporation rate, as shown in Figure 8. Additionally, warming of the atmosphere due to shortwave absorption from haze leads to less atmospheric condensation. The combined effect results in an increased specific humidity in the atmosphere, as seen from Figure 9 which shows the vertical distribution of specific humidity in both the noAER and AER cases and their differences (AER - noAER) when $CH_4/CO_2 = 0.1$. The increased specific humidity also warms up the atmosphere due to the greenhouse effect of water vapour. The overall effect of the above alters the distribution of relative humidity, as seen in Figure 10, which in turn changes the region of cloud formation. Figure 11 shows the cloud area fraction and demonstrates that the AER case has fewer clouds in the lower atmosphere compared to the noAER case. This can also be seen from Figure 12 which shows the distribution of condensed liquid water and ice clouds. This leads to a drop of cloud albedo and less shortwave radiation is reflected back to space. This top-of-atmosphere and surface cloud radiative effect is shown in Figure 13. The combined effect of clouds and haze on the overall planetary albedo is also shown in Figure 14 where the albedo for the AER case, which is ∼0.18, is lower than that for the noAER case, which is ∼0.26. The absence of higher clouds also reduces the cloud greenhouse effect, therefore decreasing the amount of longwave radiation trapped in the atmosphere. The reduction in clouds overall leads to a positive top-of-atmosphere cloud radiative effect change of 16.7 W m$^{-2}$, which is a further contribution to the increase in the global mean surface temperature by $\sim$10.6 K.

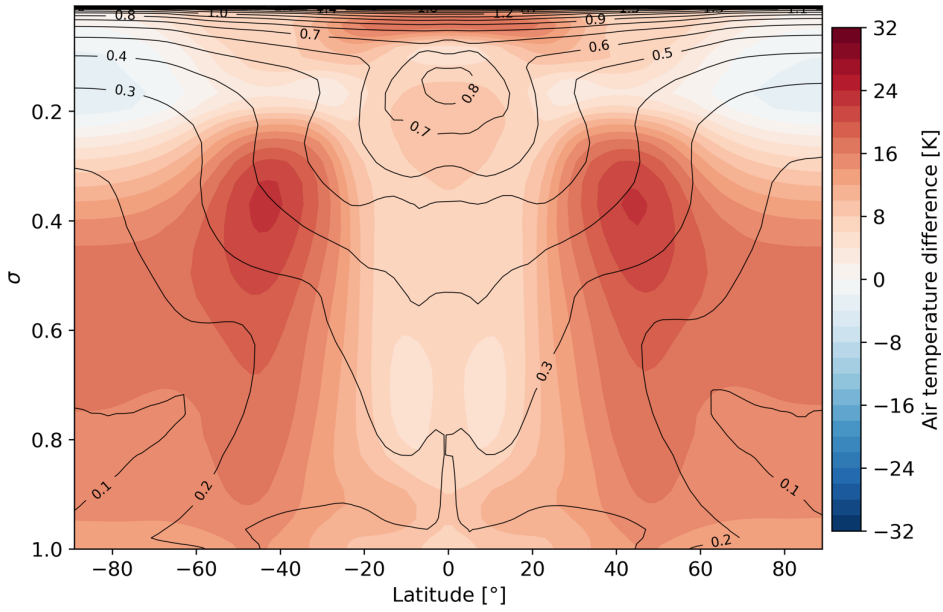

**Figure 6.** Differences in the air temperature in K between the cases with and without haze (AER - noAER) when $CH_4/CO_2 = 0.1$. The corresponding differences of zonal mean shortwave heating rate in $K\,day^{-1}$ are shown in contour lines. $\sigma$ (the y-axis) is the air pressure divided by the pressure at the surface.

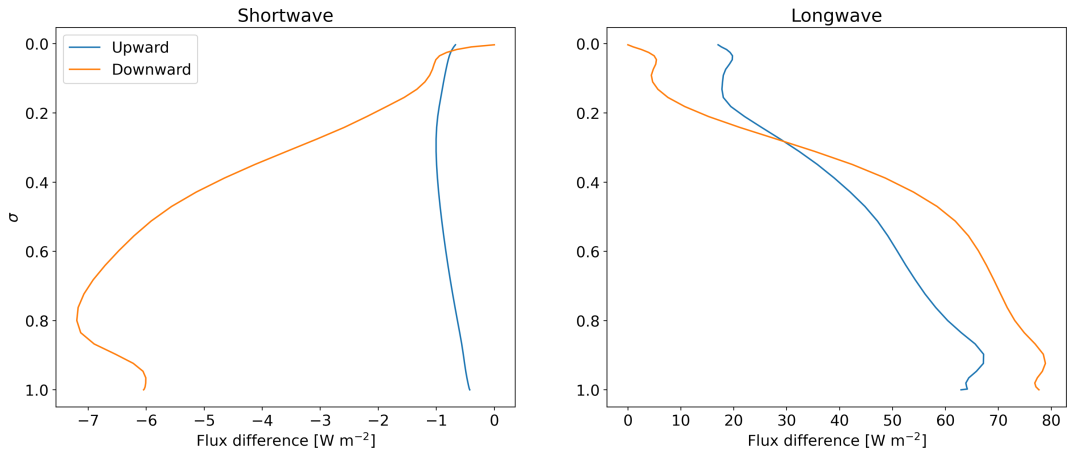

**Figure 7.** Left: Differences in the global mean cloud-free shortwave flux between the cases with and without haze (AER - noAER) when $CH_4/CO_2 = 0.1$. Right: Differences in the global mean cloud-free longwave flux between the cases with and without haze (AER - noAER) when $CH_4/CO_2 = 0.1$. $\sigma$ is the air pressure divided by the pressure at the surface.

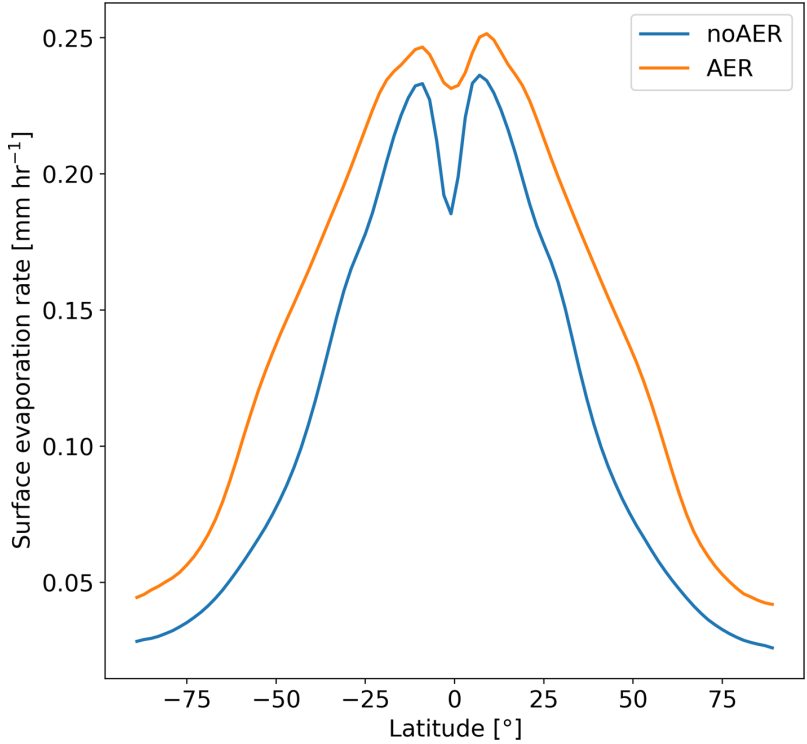

**Figure 8.** Zonal mean surface evaporation rate of noAER and AER for $CH_4/CO_2 = 0.1$.

We also look at the convection strength in the atmosphere in the presence of haze. Figure 15 shows the meridional circulation strength (see equation 4 in J. Haqq-Misra & Kopparapu, 2015; Vallis, 2017) in both the noAER and AER cases, with the contours showing the zonal mean eastward wind. The AER case demonstrates a weaker circulation strength. This is due to the increased shortwave heating in the troposphere (see contours in Figure 6), increasing the stability of the troposphere against convection, thereby weakening the meridional circulation (Eager-Nash et al., 2023). The weakened circulation reduces cloud formation, which has led to an overall increase in the total cloud radiative effect (discussed above), therefore warming up the planet. The strength of eastward wind is also weakened near the tropopause caused by the weakened circulation, especially at $\sim$30°latitude.

### 3.3 $CH_4/CO_2 > 0.1$: Haze Cooling Effect

Drastic cooling occurs for simulations including haze where $CH_4/CO_2 > 0.1$ (see Figure 3). Figure 16 shows the air temperature difference between the matching cases with and without haze (AER - noAER) when $CH_4/CO_2 > 0.1$. The largest warming occurs at the top of the atmosphere and the largest cooling occurs in the lower atmosphere at the equator. Figure 17 shows the differences in shortwave heating in K day$^{-1}$ between the matching cases with and without haze (AER - noAER) where $CH_4/CO_2 > 0.1$. This figure only shows the model levels above $\sigma = 0.08$ as these are where the shortwave heating differs the most. The heating increases as the $CH_4/CO_2$ ratio increases. The heating goes up to more than 80 K day$^{-1}$ when $CH_4/CO_2 = 0.6$. This analysis shows that

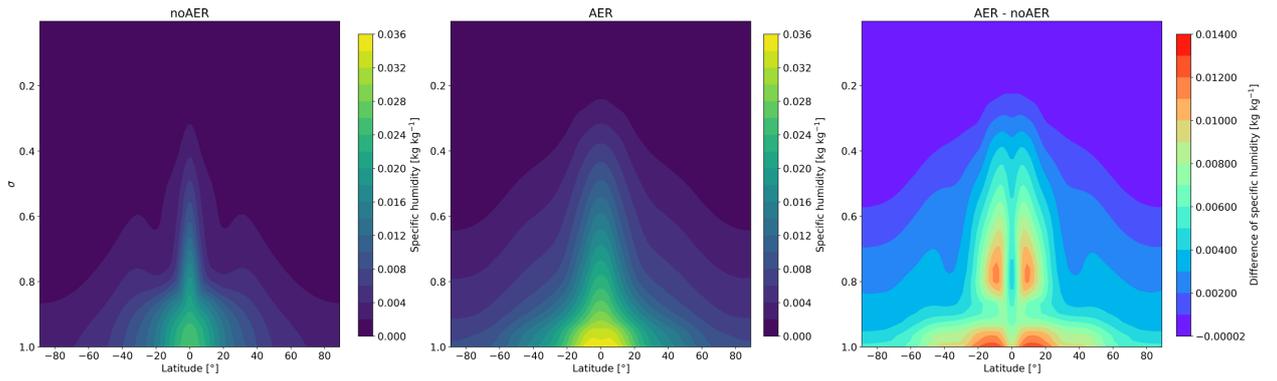

**Figure 9.** Specific humidity of noAER (left) and AER (middle) and their differences (left) for $CH_4/CO_2 = 0.1$. $\sigma$ is the air pressure divided by the pressure at the surface.

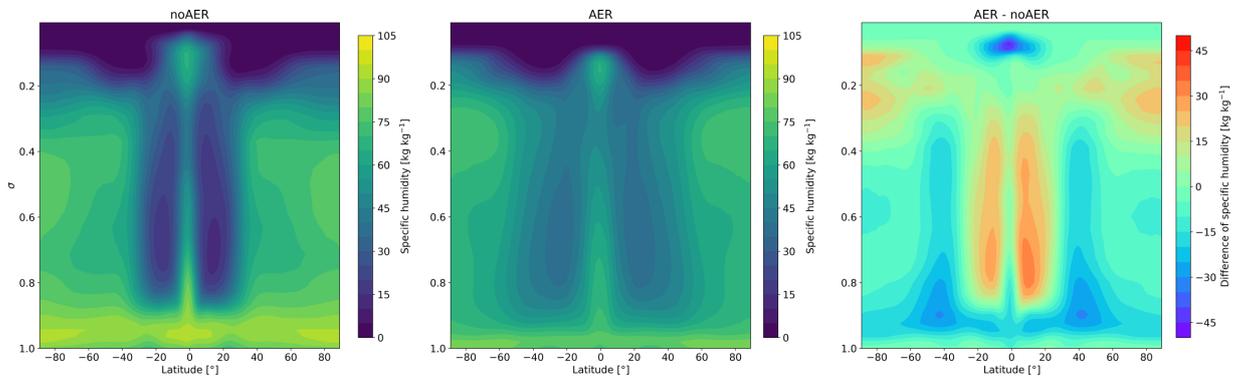

**Figure 10.** Relative humidity of noAER (left) and AER (middle) and their differences (left) for $CH_4/CO_2 = 0.1$. $\sigma$ is the air pressure divided by the pressure at the surface.

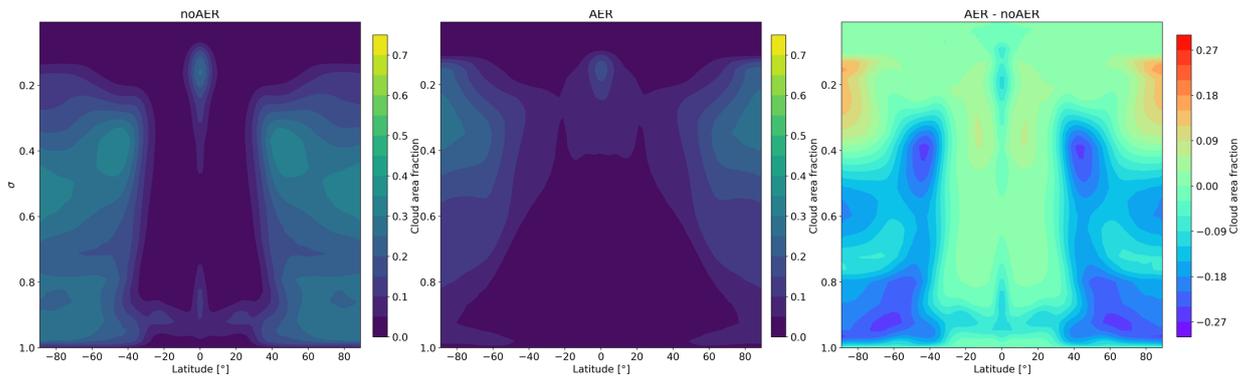

**Figure 11.** Cloud area fraction of noAER (left), AER (middle) and their differences (right) for $CH_4/CO_2 = 0.1$. $\sigma$ is the air pressure divided by the pressure at the surface.

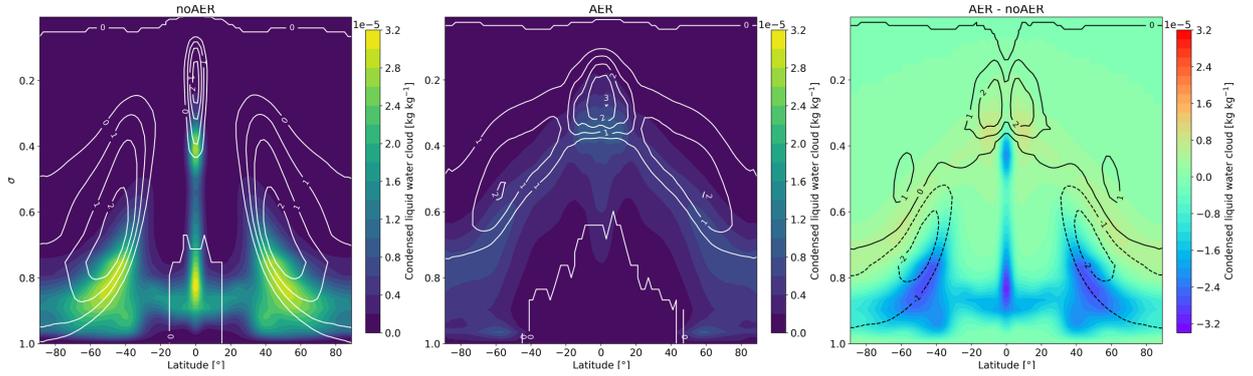

**Figure 12.** Condensed liquid water cloud of noAER (left), AER (middle) and their differences (right) for $CH_4/CO_2 = 0.1$. Condensed ice cloud of the corresponding cases with the same unit are also plotted in contour lines ($\times 10^{-5}$). $\sigma$ is the air pressure divided by the pressure at the surface.

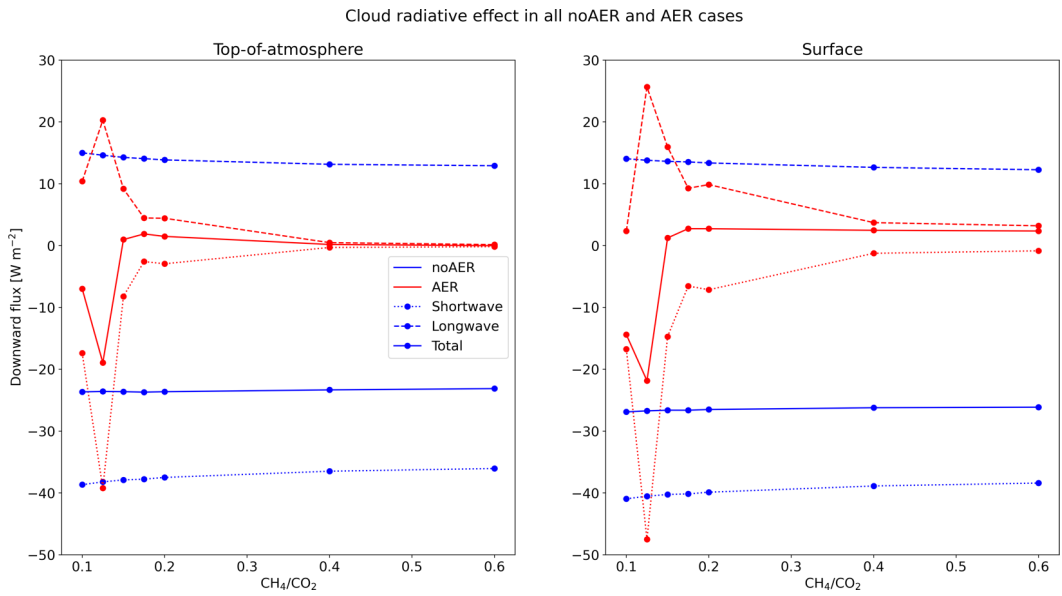

**Figure 13.** Top-of-atmosphere (left) and surface (right) cloud radiative effect of all noAER and AER cases.

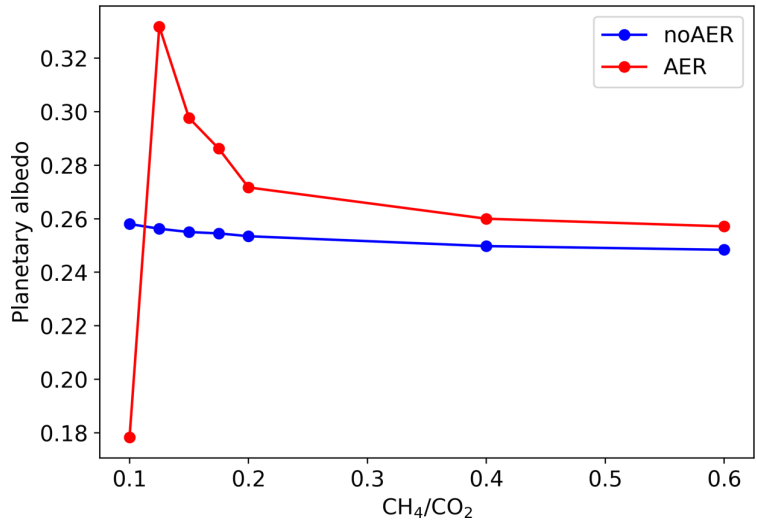

**Figure 14.** Planetary albedo off all noAER and AER cases.

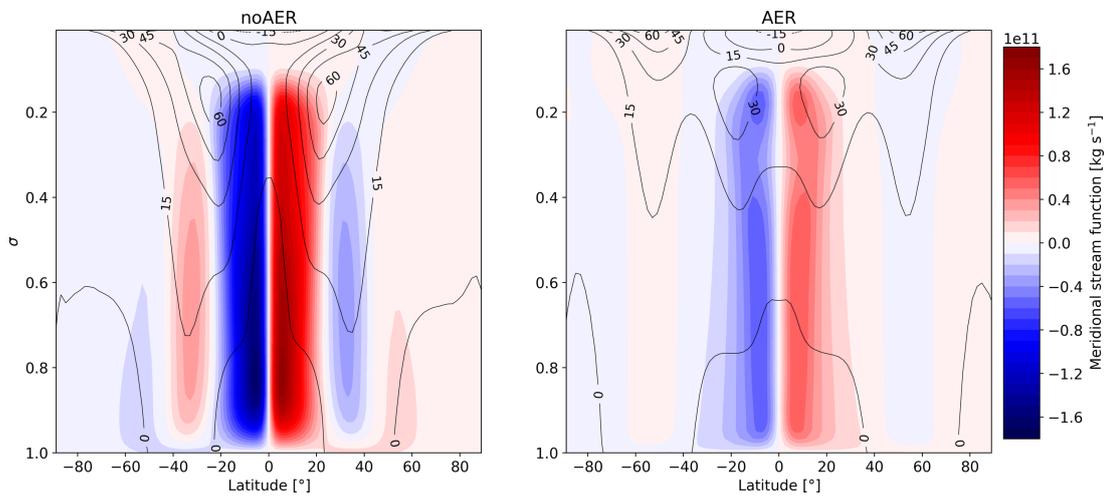

**Figure 15.** Meridional stream function, highlighting the strength of the meridional circulation of noAER (left) and AER (right) for $CH_4/CO_2 = 0.1$. $\sigma$ is the air pressure divided by the pressure at the surface.

the absorption of the shortwave radiation due to haze is the strongest around the top of the atmosphere and increases when the $CH_4/CO_2$ ratio increases. This subsequently reduces the amount of flux that reaches the deeper atmosphere, and this can be seen in Figure 18. The left panel of Figure 18 shows the difference in the global mean clear-sky shortwave flux between the matching cases with and without haze (AER - noAER) when $CH_4/CO_2 > 0.1$. This is in the absence of clouds but includes the radiative effect of haze. At the top of the atmosphere, almost all cases share an increased upward shortwave flux of $\sim 40\,W\,m^{-2}$, compared to the matching haze-free cases, except when $CH_4/CO_2 = 0.125$ where the increased scattered flux is $\sim 20\,W\,m^{-2}$, compared to its matching haze-free case. This shows that the scattering effect of the haze reaches its maximum when $CH_4/CO_2 \geq 0.15$. As the haze layer gets thicker, the amount of downward shortwave flux reaching the lower atmosphere reduces. The reflected flux due to surface albedo will again experience scattering and absorption, further reducing the amount of both upward and downward shortwave flux seen in each layer. Therefore in Figure 18, we can see a reducing amount of shortwave flux as it goes from the lower to the upper atmosphere. The reduction of downward shortwave flux is also far greater than that of the upward flux, again showing the strong absorption and scattering of haze that shortwave flux is prevented from reaching the surface. The right panel of Figure 18 shows the longwave flux difference. The upper atmosphere is exhibiting an increase of longwave flux, due to the re-emission from haze.

When the haze layer becomes thicker, the atmosphere becomes colder. This reduces the surface evaporation rate but increases the relative humidity in the atmosphere. But due to a lower tropopause (see Table 1), the specific and relative humidity is confined to the lower atmosphere, as seen from Figure 19. The noAER cases are not included here as they all share a similar distribution to that shown in Figure 10 for the $CH_4/CO_2 = 0.1$ case. This further alters the cloud distribution which is shown in Figure 20. Figure 13 shows the top-of-atmosphere and surface cloud radiative effect for these cooling cases. Note that the case when $CH_4/CO_2 = 0.125$ actually shows an increase in abundance of lower clouds, compared to when $CH_4/CO_2 = 0.1$. The increased cloud albedo due to the increased cloud area fraction results in more longwave flux being trapped due to the cloud greenhouse effect, compared to the values in the case $CH_4/CO_2 = 0.1$ (see Figure 13). When $CH_4/CO_2 > 0.125$, the clouds are concentrated in the lower atmosphere and are almost all composed of ice due to the low temperatures. The reduction of cloud abundance leads to a drop in the cloud albedo and also weakens its greenhouse effect, therefore reducing the amount of longwave flux trapped in the atmosphere. The combined effect of clouds and haze on the planetary albedo compared to the noAER cases can again be seen from Figure 14. As discussed above, since there is an increased abundance of lower clouds when $CH_4/CO_2 = 0.125$, its albedo is a lot higher than the noAER and other AER cases. However even though there is a drop of cloud albedo for the AER cases when $CH_4/CO_2 > 0.125$, the thick layer of haze increase the planetary albedo in general. Therefore all the AER cases when $CH_4/CO_2 > 0.1$ show a drop of surface temperature.

Similar to the case when $CH_4/CO_2 = 0.1$, the addition of haze reduces the strength of the meridional circulation. Figure 21 shows the meridional stream function, indicating the strength of the meridional circulation for all the AER cases where $CH_4/CO_2 > 0.1$. The noAER cases are not included in Figure 21 here as they all share a similar strength to that shown in Figure 15 for the $CH_4/CO_2 = 0.1$ case. Figure 21 shows that as the $CH_4/CO_2$ ratio increases, the size of the circulation cell reduces in vertical range in the experiments with haze. This is due to the enhanced shortwave heating in the upper atmosphere, which results in the temperature inversion occurring lower in the atmosphere when the haze layer is thicker, as shown in Figure 3. This reduces the vertical extent of the updrafts. Moreover, as the ratio increases, the circulation cells reduce from three distinct cells ($CH_4/CO_2 = 0.125$) to two in each hemisphere ($CH_4/CO_2 > 0.125$). Figure 3 shows that the temperature difference between the tropopause and the surface reduces when the ratio increases. The driving of upward flows either through convection or large-scale flow is reduced. Balanced against the rotation, a reduction in the circulation strength eventually leads to a reduction in the number of cells. The strength of eastward wind continues to weaken around the latitude of $\sim 30°$ owing to the weakened meridional circulation. The strongest jet also appears deeper in the atmosphere, along with the decrease of size of the circulation cell.

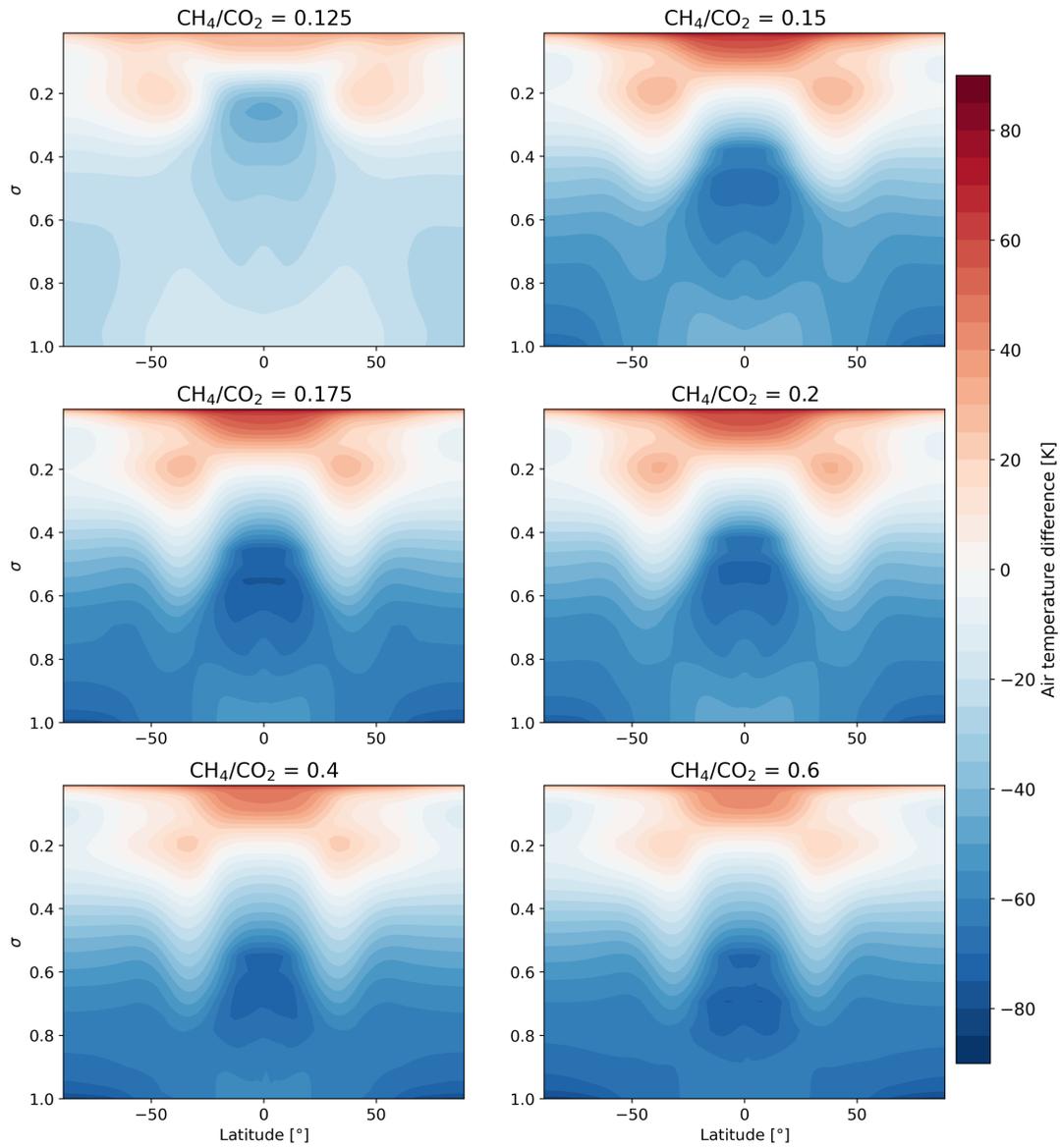

**Figure 16.** Air temperature difference in K between the matching cases with and without haze (AER - noAER) when $CH_4/CO_2 > 0.1$. $\sigma$ is the air pressure divided by the pressure at the surface.

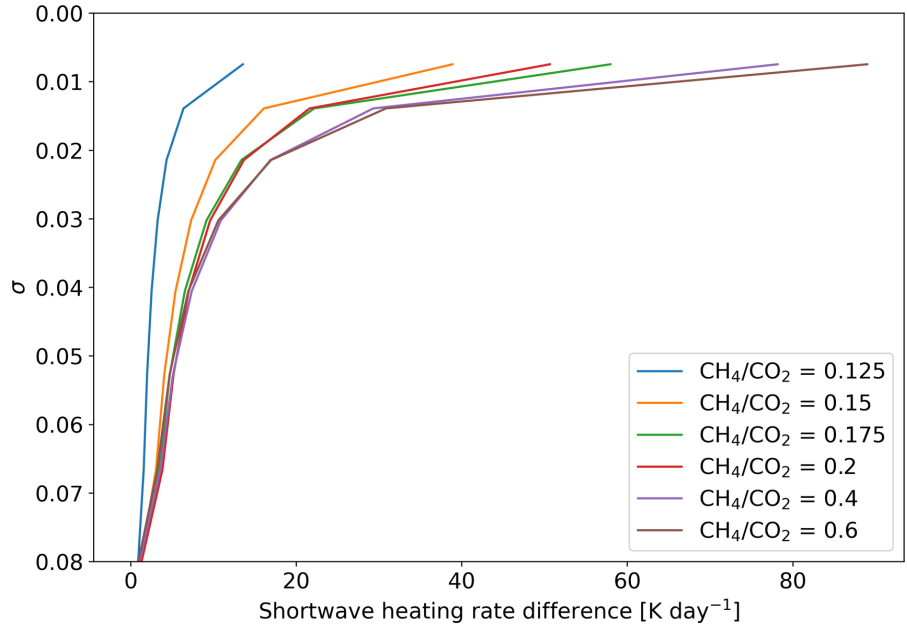

**Figure 17.** Differences in the global mean shortwave heating rate between the matching cases with and without haze (AER - noAER) when $CH_4/CO_2 = 0.1$ at the levels above $\sigma = 0.08$. $\sigma$ is the air pressure divided by the pressure at the surface.

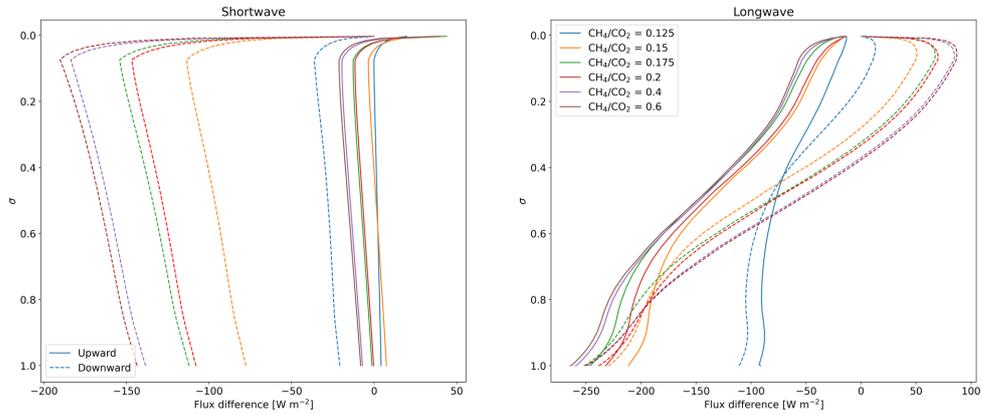

**Figure 18.** Left: Differences in the global mean cloud-free upward shortwave scattered flux between the matching cases with and without haze (AER - noAER) when $CH_4/CO_2 > 0.1$. Right: Differences in the global mean cloud-free downward shortwave scattered flux between the matching cases with and without haze (AER - noAER) when $CH_4/CO_2 > 0.1$. $\sigma$ is the air pressure divided by the pressure at the surface.

The results from Section 3.2 and Section 3.3 are summarised in Figure 22. On the left, in the absence of haze, there is a substantial amount of lower and high clouds, which facilitate the reflectance of shortwave due to the cloud albedo effect. High cloud also helps to warm up the planet via the greenhouse effect. In the middle panel, for a thin layer of haze, scattering and absorption occurs and warms up the atmosphere. This increases the specific humidity in the atmosphere and warm up the atmosphere due to the greenhouse effect of water vapour. This also alters the evaporation rate and relative humidity, which then inhibits the formation of clouds in the lower atmosphere. A reduced abundance of clouds lessens the cloud albedo effect and overall allows more shortwave radiation reaching the surface. Therefore we see an increase of $\sim 10.6$ K in the global mean surface temperature. On the right, for a case of a thick layer of haze, more shortwave is being scattered and absorbed in the upper atmosphere, and less flux is able to reach the surface. This results in a drastic cooling and the formation of clouds in the lower atmosphere.

# 4  Discussion

In this work, we have shown how the presence of non-interactive spherical hydrocarbon haze changes the energy budget of the planet, which drives changes, relative to a commensurate haze-free simulation, in the meridional mean circulation and cloud formation. The resulting changes in the cloud radiative feedback act to further warm or cool the planet. As detailed in Section 3.2, we found that when $CH_4/CO_2 < 0.15$, the simulations are able to support liquid water on the surface, therefore fulfilling the habitable condition. For ratios larger than that, not only does the thick layer of haze produce uninhabitable surface conditions, but it is also unlikely to persist in the atmosphere of the Archean Earth as the biogenic $CH_4$ production would be limited when the surface temperature becomes too cold. J. D. Haqq-Misra et al. (2008) have also suggested that the Archean $CH_4$ concentration should be stabilized under the presence of a thin layer of haze. As a result, the Archean Earth is very likely to have been enveloped in only a thin layer of haze, at most.

However, again note that our work here is under the setup of a non-interactive spherical haze scheme. The results might differ from a 3D model with a self-consistent fractal haze formation. In the following, we compare our work with previous 1D results, followed by detailing the caveats, and future work, relevant to our study.

## 4.1  Comparison With Previous 1D Results

The use of 3D models in our study suggests that a thin haze layer can lead to a substantial increase in the global mean surface temperature, which could contribute to resolving the FYS problem (Sagan & Mullen, 1972). This contrasts to previous 1D studies none of which have shown an increase of surface temperature due to the presence of haze. They have shown that haze either has no radiative impact on the Archean climate, or that it would act as a barrier preventing shortwave radiation from entering the atmosphere due to its shortwave absorption property, which ultimately cools the atmosphere. Our work shows that the interaction between the haze and the general circulation plays an important role in decreasing cloud coverage, which contributes to global mean temperature increase of $\sim 10.6$ K. The haze layer at $CH_4/CO_2 = 0.1$ could lower the $CO_2$ concentrations required to keep the Earth out of a snowball state, and potentially be part of a solution to the FYS problem.

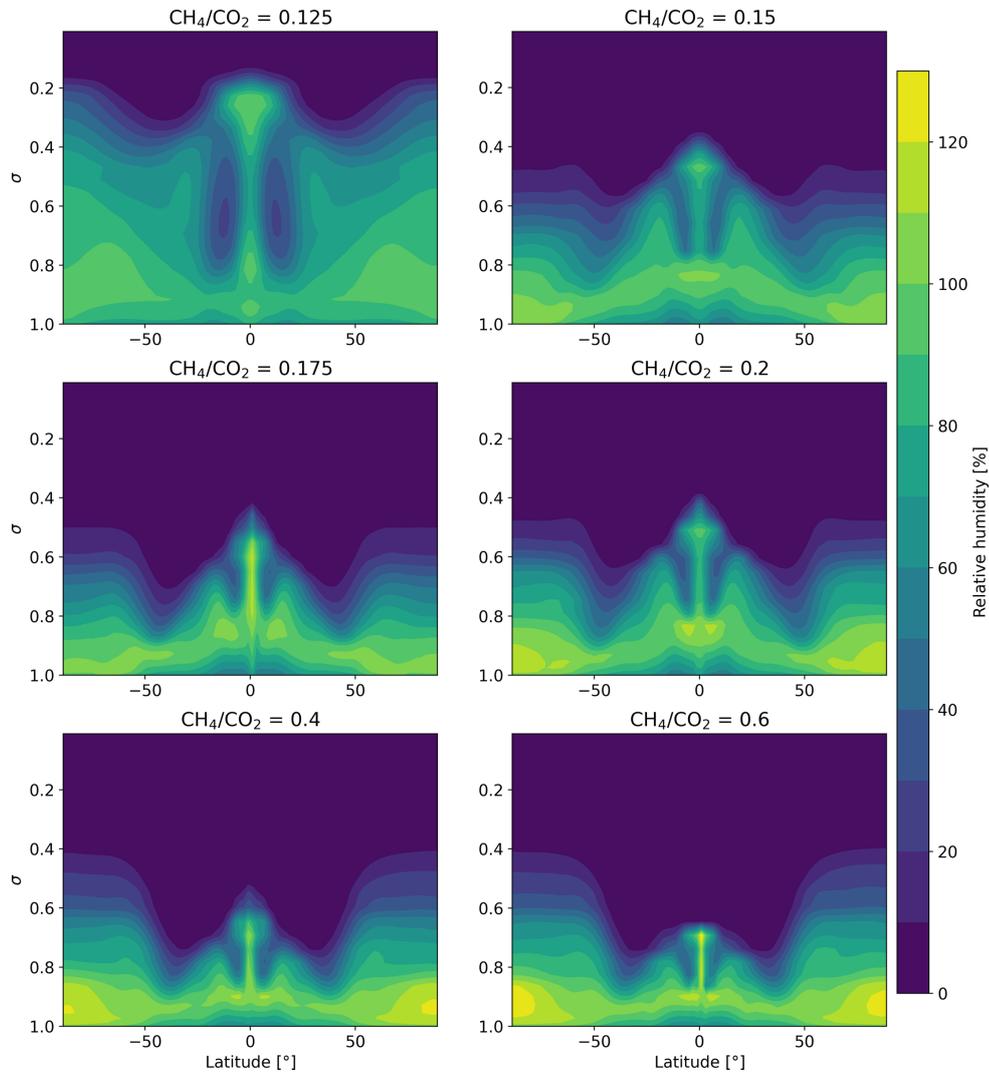

**Figure 19.** Relative humidity in the AER cases when $CH_4/CO_2 > 0.1$. $\sigma$ is the air pressure divided by the pressure at the surface.

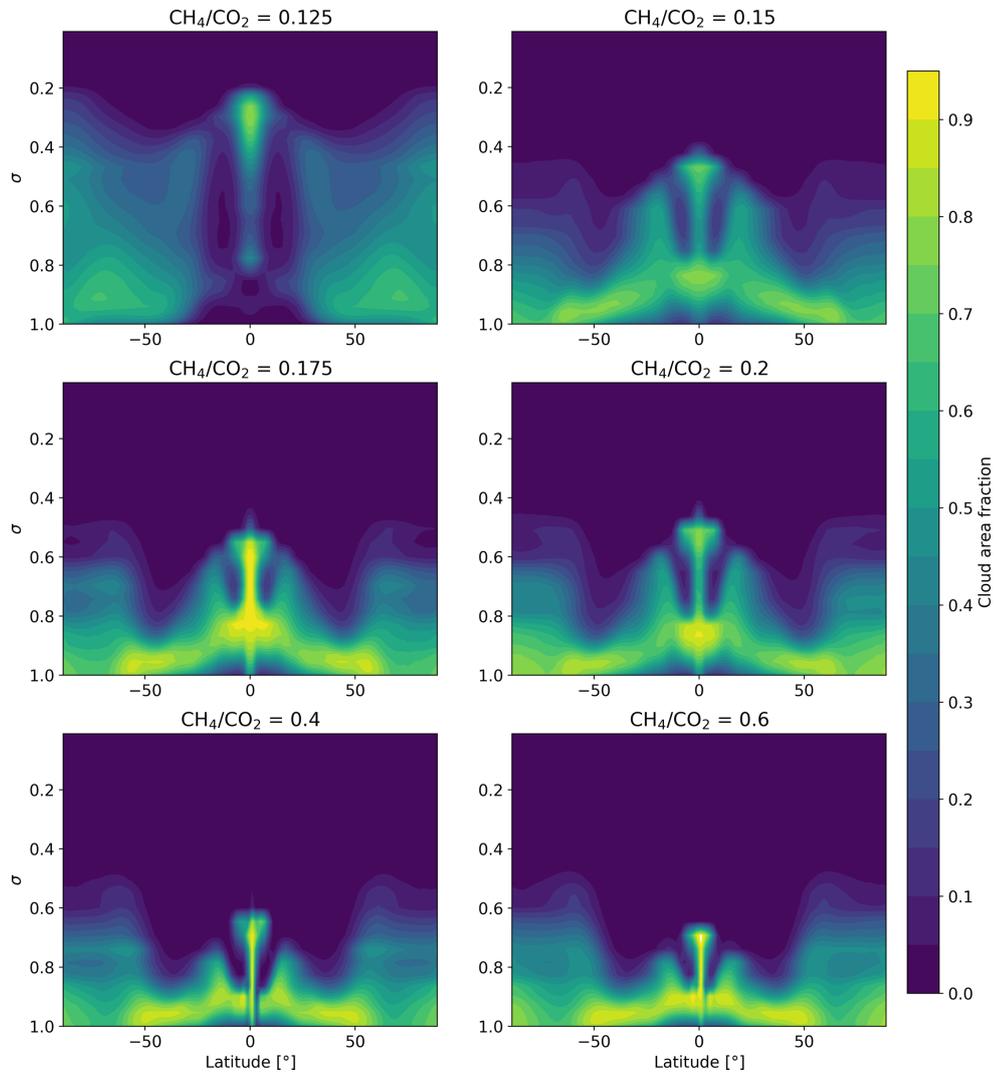

**Figure 20.** Cloud area fraction of AER when $CH_4/CO_2 > 0.1$. $\sigma$ is the air pressure divided by the pressure at the surface.

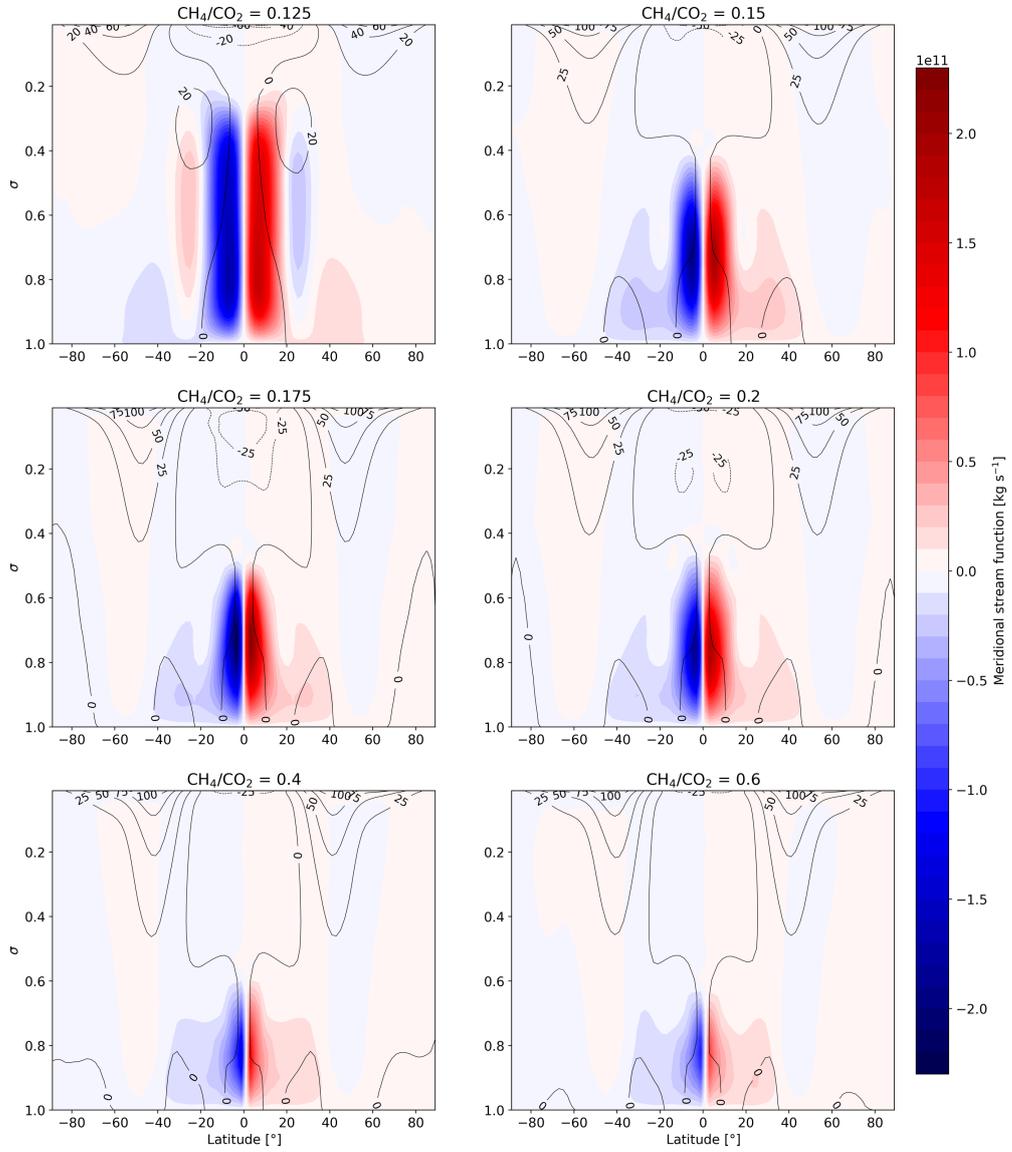

**Figure 21.** Mean meridional mass stream function in the AER cases when $CH_4/CO_2 > 0.1$. $\sigma$ is the air pressure divided by the pressure at the surface.

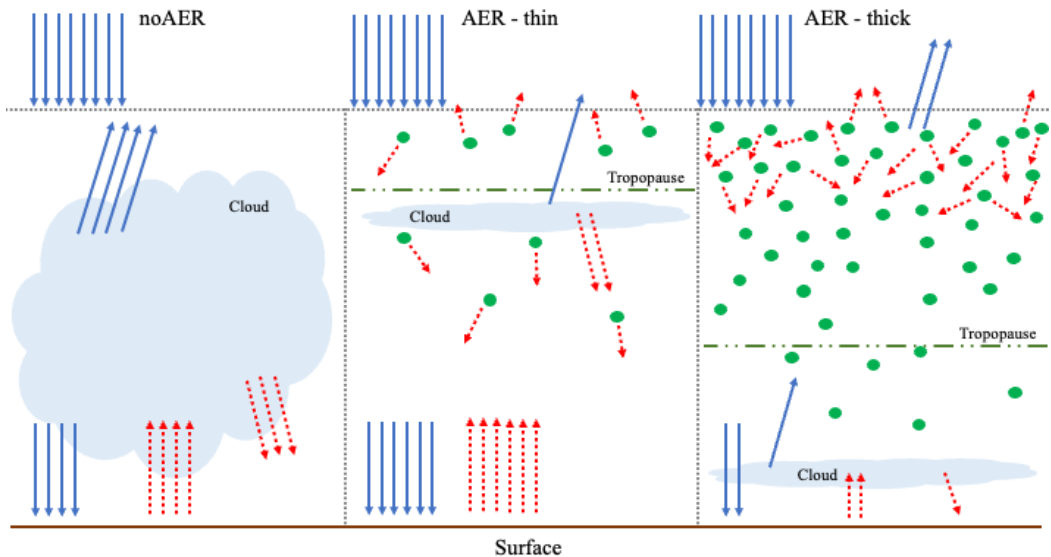

**Figure 22.** Schematic of the changes of shortwave and longwave radiation in noAER case (left) and in AER cases for thin (middle) and thick haze (right). Blue solid arrows represent shortwave radiation and red dashed arrows represent longwave radiation. Number of arrows shows the strength of radiation. Green dots represent haze.

For a larger $CH_4/CO_2$ ratio, our results share the same trend with those found in 1D. When the haze layer becomes thicker and the shortwave scattering and absorption in the upper atmosphere dominates, the planet will cool down rapidly regardless of the ratio. The stronger thermal inversion and a lower tropopause inhibit the overturning circulation. The low temperature in the lower atmosphere facilitates the formation of ice clouds over liquid clouds. Similar to Arney et al. (2016), we find that when the haze reaches a threshold optical thickness, further increases in the amount of haze (i.e. increasing $CH_4/CO_2$ ratio) would only lead to a plateau in the surface temperature (see Figure 4 when $CH_4/CO_2$ = 0.4 and 0.6).

To summarize, our work shows a different climate behavior due to haze in a 3D model compared with the previous 1D results. This shows the importance of using a 3D model to capture the effect of haze in the Archean climate. However, in this initial study we have still made simplify assumptions that will be addressed in a future work. Starting from a more idealised treatment, and gradually increasing the complexity of the modelling approach is vital to build up an understanding of the behaviour of complex systems such as the climate of the Archean Earth.

### 4.2 Future Work

In this work, regarding the atmospheric composition, we have only considered $N_2$, $CO_2$ and $CH_4$ in our model setup. However, as J. D. Haqq-Misra et al. (2008) have pointed out, $C_2H_6$ might also be formed due to the high concentration of $CH_4$ which could then act to warm up the Archean climate. If our work included $C_2H_6$, we might expect a stronger warming when $CH_4/CO_2$ = 0.1 and that the haze warming effect might also occur slightly above the ratio of 0.1. Regarding our haze treatment, as noted in Arney et al. (2016), the simplified haze chemical scheme may under-estimate haze production because the particles will grow much larger before they condense as aerosols, and back-reactions may occur. As discussed in Section 2.2, we consider the refractive indices of the Titan-like haze analogue from Khare et al. (1984) and He et al. (2022). However, as briefly discussed in Section 1, the hydrocarbon haze produced in Titan would be different from that produced in the Archean Earth due to the different atmospheric composition. For example, Mahjoub et al. (2012) studied the optical constants produced by different $CH_4$ concentrations in a mixture of $N_2$ and $CH_4$ and they found that the optical properties depend strongly on the $CH_4$ concentration. Arney et al. (2016) has also pointed out that their results strongly depend on the choice of optical properties of the haze. Therefore our results might not be a completely accurate representation of the effect of the specific haze present during the Archean. Furthermore, our model domain only extends to 39.25 km but according to the output from Atmos, the haze number density distribution peaks above $\sim$70 km. As a result, future work is needed to examine all the effects of haze in the upper atmosphere. However, similar to Wolf and Toon (2010), the haze opacity peaks within the model domain of our work here. Additionally, we have applied Mie scattering to calculate the extinction coefficient of the haze particles. However, studies have been shown that hydrocarbon particles are actually fractal shapes (spherical monomers sticking together to form aggregates), rather than spheres (Bar-Nun et al., 1988). Wolf and Toon (2010) compared the optical properties between these two types and show that spherical particles exhibit a constant extinction strength from the ultraviolet (UV) to the visible regime, while fractal particles lead to more extinction in the UV but remain optically thin at visible wavelengths compared to spherical haze. The solar spectrum used in this work peaks at $\sim$0.45 $\mu$m. Therefore, our use of Mie scattering and the underlying assumption of spherical haze particles may overestimate the surface cooling and atmospheric heating compared to fractal haze. In future work we will extend the model domain and incorporate the mean-field approximation of Mie scattering (Botet et al., 1997), as adopted in Zerkle et al. (2012) and Arney et al. (2016), and take into account the optical properties of fractal haze.

Also, an important caveat regarding the photochemical modeling is that the two models used here are not fully coupled, so the temperature from the climate model is not iteratively fed back to the photochemical model. An interactive haze scheme will be able to demonstrate the presence of the negative feedback loop where haze production from methane photolysis is effectively shut down once the haze layer becomes too thick and shields methane from UV radiation, as found by both Zerkle et al. (2012) and Arney et al. (2016). It can also capture the changes in haze production due to the change of surface temperature and methane production. The self-consistent scheme can further capture the temperature feedbacks on the balance between the particle coagulation timescale and the timescale for removal (Arney et al., 2016). For instance, the stratospheric warming effects of hazes might reduce the particle coagulation timescale, allowing particles to move faster and to collide more frequently. The haze particle sizes might increase, as seen from the work of Arney et al. (2016). In this work, the photochemical model is initiated with a haze-free temperature profile from the climate model, and the temperature feedbacks are not explored. The results here might underestimate the haze particle sizes. The 3D model will also capture the advection of haze particles away from the photochemical production sites. For example, Wolf and Toon (2010) have shown that in their 3D model, the haze layer, which is radiatively inactive, is thinner in the equatorial region as the haze particles are transported to the polar region. Even though they have suggested that in the haze's radiatively active state this latitudinal gradient might be reduced, we speculate that this might still weaken the impact of haze in terms of its direct radiative effect. Furthermore, if the already thin haze layer (considering the case when $CH_4/CO_2 = 0.1$) is even thinner near the equator, the cloud area fraction might increase as the atmosphere is less warmed up by the haze (compared to our experiments). As a result, the planetary albedo might increase. However for an already thick haze layer, the haze transport from equator to pole might not have much effect. Last but not least, if the haze is coupled with the cloud schemes (which is not implemented in this work), haze particles could serve as cloud condensation nuclei and increase the cloud optical depth.

The cloud albedo would increase. At the same time, haze could also be removed as rain out which might cancel out this effect. All these might change the impact of haze on the climate found in this work and reduce its warming or cooling effect, demanding further study.

## 5 Conclusions

We prescribe a fixed spherical haze profile within a 3D GCM, the Met Office Unified Model, to understand the impact of haze on the atmosphere of the Archean Earth. The haze profiles were generated using an input temperature pressure profile from a haze-free simulation of the Archean (Eager-Nash et al., 2023) and the photochemical model Atmos, for a range of $CH_4/CO_2$ ratios. Fixing $pCO_2$ at 3,000 Pa and varying the $pCH_4$, we found that when $CH_4/CO_2 = 0.1$, the warming goes up to $\sim 10.6$ K when compared with the simulation without haze, unlike previous 1D work that show either cooling or no obvious temperature change in such a case. Our work shows that the temperature change for a thin haze layer is determined by the water vapour and the cloud feedback due to the warming in the upper atmosphere. However, a thick layer of haze when $CH_4/CO_2 > 0.1$ leads to cooling as the shortwave radiation is being absorbed or scattered in the upper atmosphere instead of being absorbed by the surface. As a result, a temperature inversion is formed in the upper atmosphere, while the surface temperature is reduced drastically. Even so, by comparing the benchmark temperature of 273 K as a habitable condition, our results show that habitable regions exist up to when $CH_4/CO_2 = 0.125$. More work is needed to couple the interactive haze to the climate model to fully capture the feedback between haze formation and the atmospheric dynamics.

### Open Research

The research data supporting this publication are openly available with CC BY 4.0 (Mak et al., 2023).

### Acknowledgments

We acknowledge funding from the Bell Burnell Graduate Scholarship Fund, administered and managed by the Institute of Physics, which made this work possible. This work was supported by a UKRI Future Leaders Fellowship [grant number MR/T040866/1], a Science and Technology Facilities Council Consolidated Grant [ST/R000395/1] and the Leverhulme Trust through a research project grant [RPG-2020-82]. Material produced using Met Office Software. We acknowledge use of the Monsoon2 system, a collaborative facility supplied under the Joint Weather and Climate Research Programme, a strategic partnership between the Met Office and the Natural Environment Research Council. This work used the DiRAC Complexity system, operated by the University of Leicester IT Services, which forms part of the STFC DiRAC HPC Facility (www.dirac.ac.uk). This equipment is funded by BIS National E-Infrastructure capital grant ST/K000373/1 and STFC DiRAC Operations grant ST/K0003259/1. DiRAC is part of the National e-Infrastructure. G. Arney acknowledges funding from the Virtual Planetary Laboratory Team, a member of the NASA Nexus for Exoplanet System Science, funded via NASA Astrobiology Program Grant No. 80NSSC18K0829. JE-N would like to thank the Hill Family Scholarship. The Hill Family Scholarship has been generously supported by University of Exeter alumnus, and president of the University's US Foundation Graham Hill (Economic & Political Development, 1992) and other donors to the US Foundation.